# Modeling of Zircaloy Oxidation Through Dynamic Mesh Deformation

A. Scolaro[1], E. Brunetto[1] and C. Fiorina[1,2]

[1]École Polytechnique Fédérale de Lausanne (EPFL) - Laboratory for Reactor Physics and Systems Behavior (LRS), Route Cantonale 1015, Lausanne, Switzerland

[2]Texas A&M University, Department of Nuclear Engineering College Station, TX 77843-3133, USA

alessandro.scolaro@epfl.ch, edoardo.brunetto@epfl.ch, carlo.fiorina@tamu.edu

## ABSTRACT

Zircaloy cladding corrosion in Light Water Reactors (LWRs) results in the formation of an outer oxide layer and in the thinning of the metallic portion of the cladding. At the École Polytechnique Fédérale de Lausanne (EPFL), we aim to investigate the impact of asymmetric oxide or crud layers and their feedback on thermal-hydraulics. To achieve this, we are implementing mechanistic capabilities for the treatment of corrosion within the open-source multi-dimensional code OFFBEAT. In this work, we outline our approach based on the combination of traditional laws for predicting the growth of the oxide layer with methods to dynamically modify the mesh points and topology. Specifically, our methodology employs routines available in the OpenFOAM library to add or remove layers of cells according to the growth of the oxide layer, along with a mesh motion solver to maintain mesh quality. We verify the methodologys against an analytical test case and then we demonstrate its potential for the study of non-uniform oxide distributions by modeling a simplified cladding ring with asymmetric temperature profile, mimicking the conditions of azimuthal power peaking in a Boiling Water Reactor (BWR).

KEYWORDS: corrosion; dynamic mesh; layer addition-removal; OpenFOAM; OFFBEAT

## 1. INTRODUCTION

Water-side corrosion of the Zircaloy cladding used in LWRs limits the fuel rod lifespan and the extension towards higher burnups, and it represents a safety concern during accident scenarios such as LOCAs. The formation of an outer oxide layer affects both the thermal and the mechanical behavior of the fuel rod: first, the thermal conductivity of $ZrO_2$ is significantly lower than that of Zircaloy, thus even a thin oxide film might impair the heat transfer with the coolant; then, as the layer grows larger, the metallic cladding is consumed and becomes thinner, leading to changes in the stress and strain distribution; additionally, hydrogen produced during oxidation can diffuse into the metal and form hydrides causing the cladding to become more brittle.

Most fuel performance codes used today do not explicitly model the oxide layer, instead including its effect on the temperature distribution through an additional 1-D thermal resistance. While some codes can model the thinning of the metallic cladding, these capabilities have been introduced relatively recently [1] and had a limited number of applications. Modern multi-dimensional codes are interesting tools for studying non uniform oxide distributions, although modifying the location of the computational nodes/cells is less straightforward in the presence of 3-D unstructured meshes. Recently there have been some efforts in this direction for some of the more modern Finite Element Method multi-dimensional codes, such as the work done by Bailly-Salins et al. using the XFEM within the MOOSE platform [2], or the work done by Khostov [3] using the Falcon code

In this framework, at the EPFL in Switzerland, we are developing mechanistic methodologies to capture the evolution of the metal-oxide interface, the reduction of the cladding thickness and the growing oxide layer. We are implementing these models in the open-source multi-dimensional fuel performance code OFFBEAT [4] which is based on the Finite Volume Method C++ library OpenFOAM. Our final objective is to study the corrosion of the cladding under asymmetric irradiation conditions, eventually coupling thermo-mechanical analysis with thermal-hydraulic solvers available in OpenFOAM, such as the GeN-Foam platform [5].

This preliminary work outlines our methodology for modeling the corrosion of Zircaloy cladding in LWRs. Our approach involves using correlations available in the open-literature to calculate the thickness of the $ZrO_2$ layer and then dynamically modifying the mesh topology with mesh alteration routines available in OpenFOAM. To ensure mesh quality and prevent excessive distortion, we also employ a mesh motion solver that solves a diffusion equation for the mesh motion field. In the following section we provide a detailed description of the methodology. We then present a simple verification case, followed by a more complex test-case mimicking the conditions of local power peaking during a BWR dryout. This test-case highlights the potential of our methodology to accurately capture non uniform oxide layers.

## 2. METHODOLOGY

In this section, we describe the novel corrosion methodology implemented in OFFBEAT. First, we provide an overview of the key components, including the dynamic mesh routines that allow for morphing the mesh as the oxide layer grows. Then, we present two variations of the methodology, both currently available in OFFBEAT: an *implicit-oxide* approach, which models the change of dimensions of the cladding, without accounting for the oxide layer with separate cells; and an *explicit-oxide* approach, which takes advantage of the dynamic mesh capabilities to model both the reduction of the metallic cladding thickness and the growth of the oxide layer.

### 2.1. Basic elements

The methodology relies on the following three basic components:

1. **Standard correlations for predicting the growth of the oxide layer**. We rely on the correlations described in the documentation of the *ZryOxidation* class from the MOOSE framework [6], which in turn is based on previous works by Ritchie [7] and Schanz [8]. This includes considering two temperature regimes: normal operation between 523 K and 673 K, and a higher temperature regime where water turns to steam, and oxidation proceeds more rapidly. Arrhenius equations are used to capture the time-dependent growth of the

Modeling of Zircaloy Oxidation Through Dynamic Mesh Deformation

oxide layer $S_{ox}$, which is either cubic, linear or quadratic depending on the temperature regime. At each time step, from the incremental increase of oxide thickness $\Delta S_{ox}$ we derive the incremental decrement of the metallic cladding thickness $\Delta S_{met}$ using the Pilling-Bedworth ratio which is for Zirconium is equal to 1.56:

$$R_{PB} = \frac{V_{ox}}{nV_{met}} \cong \frac{A \cdot \Delta S_{ox}}{n \cdot A \cdot \Delta S_{met}} = \frac{\Delta S_{ox}}{n \cdot \Delta S_{met}} \tag{1}$$

2. **A mesh motion solver**. This solver applies a diffusion (Laplace's) equation with a specified diffusivity to calculate the cell-center values of the motion displacement field in response to a fixed displacement prescribed either at the boundaries of the domain or at *faceZones* (an OpenFOAM term used to indicate a subset of mesh faces). Our mesh motion solver is adapted from the *displacementLaplacianFvMotionSolver* class already implemented in OpenFOAM. Its purpose is to maintain mesh quality while ensuring that the displacement of mesh points at specific locations (i.e. the outer cladding surface and the metal/oxide interface) corresponds to the oxide growth and metal thinning predicted by the aforementioned correlations.
3. **Layer addition and removal routines**. We make use of the *layerAdditionRemoval* mesh modifier class already implemented in the OpenFOAM library. This class allows layers of mesh cells to be added or removed while the mesh is morphed in response to changes in the domain. A "layer" in this context refers to a set of cells that are adjacent to a boundary (or to a faceZone), and that are connected to cells from the same layer in the direction parallel to the boundary. The removal/addition algorithm is illustrated in Figure 1 and it works as follows:
   a. When the layer shrinks, the outermost layer of cells on the surface of the computational domain is removed if the new layer thickness $\Delta_{upd}$ falls below a certain removal threshold $\Delta_{rem}$.
   b. Alternatively, when the layer grows, a new layer of cells is added to the surface of the computational domain if the new layer thickness $\Delta_{upd}$ is above a certain addition threshold $\Delta_{add}$

Changes in mesh connectivity are internally handled by OpenFOAM. In our re-implementation of the layerAdditionRemoval class, the user provides the removal/addition thresholds $\Delta_{rem}$ and $\Delta_{add}$ via the main input dictionary of OFFBEAT, expressed as a fraction of the original layer thickness.

## 2.1. Implicit-oxide approach

The implicit-oxide approach models the expansion of the cladding due to the growth of the oxide layer, without explicitly considering the oxide as a separate material zone. To maintain the correct total cross-sectional area, the cladding is expanded at each time step by the difference between the incremental growth of the oxide layer $\Delta S_{ox}$ and the decrement of metallic thickness $\Delta S_{met}$. Naturally, one of the main limitations of this approach is that it assumes that the entire cladding is made of metallic zircaloy. This may not be accurate for cases with large oxide layers.

The implicit approach introduces an additional inconsistency from a thermal analysis perspective, as the outer temperature of the metallic cladding is not entirely accurate. Indeed, the model contains an additional and unrealistic small portion of metallic material (identical to $\Delta S_{ox}$). While this error does not significantly impact the thermal behavior of the cladding itself (even for large heat fluxes and oxide layers one can expect an additional temperature drop of maximum ~10 degrees, which is much lower than the further jump in the oxide layer), nonetheless it must be carefully considered. The first issue is that the temperature at the outer boundary does not correspond to the outer oxide temperature. To resolve this, a dedicated boundary condition is used to consider the additional 1-D thermal resistance due to the oxide layer. This ensures that the outer temperature is accurate and the heat flux at the boundary is correct. Furthermore, since there is no explicit oxide layer in the model, the metal/oxide interface required by the corrosion correlations is not directly available. This interface temperature can be simply retrieved by assuming 1-D heat transfer between the interface, the boundary of the model, and the non-modeled oxide layer.

The proposed algorithm for the implicit approach consists of the following steps:

1. $\Delta S_{ox}$ **and** $\Delta S_{met}$ **are calculated using the procedure described in Section 2.1**. This step is performed during each outer iteration of the main OFFBEAT solve.
2. **The mesh motion solver is employed to calculate the mesh deformation field.** This and the following steps take place at the beginning of the following time step. Since OpenFOAM is a cell-centered code, the mesh deformation due to the oxide layer growth is solved first for the cell centers, and then interpolated back to the mesh points for mesh movement. At each face of the outer cladding boundary, the local net increment $\Delta S_{net} = (\Delta S_{ox} - \Delta S_{met})$ calculated at the end of the previous time step is transformed into an outward incremental displacement with magnitude $\Delta S_{net}$ and normal to the surface itself. This deformation is then prescribed as a boundary condition for the mesh motion solver, which solves Laplace's equation and obtains the cell-center motion displacement field for the rest of the mesh.
3. **Mesh update**. The cell-center motion displacement field is interpolated to the mesh points, which are moved to their new positions. The mesh deformation discussed in this work is stress-free as it is not included in the calculation of the cladding strain or stress field.
4. **Check mesh for layer removal**. Once the mesh is updated, the thickness of the outer cladding cell layer is checked. If it is found that in correspondence of at least one outer face, the cell thickness is above the addition threshold, then a new layer is added as shown in Figure 1. Finally, the OFFBEAT solve moves to the next time step.

The main benefit of the implicit approach is its simplicity and numerical efficiency. By not modeling the oxide layer as a separate zone or layer of cells, it allows for the use of coarser meshes and time steps without sacrificing accuracy. However, this approach is limited by the assumption of 1-D heat transfer in the oxide layer, which may not hold true for non-uniform thick oxide layers. Moreover, ignoring the presence of the oxide layer in the calculation of the cladding stress distribution becomes less realistic as the layer grows thicker.

Additionally, it is worth noting that the algorithm described above (as well as the one for the explicit approach) is explicit in time, meaning that the expansion of the cladding lags always one step behind and is not fully converged in a single time step. While this will be improved upon by future efforts, we expect that due to the gradual formation of the oxide layer in time the error introduced by this approach is negligible, provided that the time step size is appropriately chosen.

## 2.2. Explicit-oxide approach

The explicit-oxide approach models the growth of the oxide layer by explicitly creating a new material zone, while tracking the thinning of the metallic cladding to maintain the correct cross-sectional area. In this approach, the outer oxide temperature $T_{ox,o}$ coincides with the temperature at the outer boundary of the domain, while the interface temperature $T_I$ is automatically calculated by the thermal solver. The explicit approach offers a more accurate modeling of the cladding behavior, particularly under extreme conditions where the oxide layer may significantly affect the thermal and mechanical response of the cladding.

The algorithm for the explicit-oxide approach proceeds as follows:

1. **Calculate $\Delta S_{ox}$ and $\Delta S_{met}$ as explained in Section 2.1** through the outer iterations.
2. **The mesh motion solver is employed to calculate the mesh deformation field** (only at the beginning of the following time step). As for the implicit approach, the net increment $\Delta S_{net}$ is applied at the outer boundary of the cladding and the corresponding outward displacement is prescribed as a boundary condition for the mesh motion solver. If the oxide material zone and oxide/metal interface faceZone have already been created (as described in step 4.a), an inward displacement equal to $\Delta S_{met}$ is applied to oxide/metal interface. Then, the mesh motion solver is used to calculate the cell-center motion displacement field.
3. **Mesh update** (as described in Section 2.2).
4. **Cell layer removal and cell layer addition**. The thickness of the outer layer is checked and if this exceeds the addition threshold $\Delta_{add}$, a new layer is generated. Then,
    a. When a new layer is added for the first time, the new cells are assigned to a different *material zone* with material properties corresponding to those of the Zirconium oxide (the outer cladding boundary and relative boundary conditions, as well as its layerAdditionRemoval modifier, are now connected to the outer side of the oxide layer). Also, the faces separating the new oxide zone from the metallic cladding are collected into an *interface* faceZone and layerAdditionRemoval modifiers are applied to the newly created interface faceZone in order to check the layer thickness on both side of the interface. From now on, the mesh update steps 2-3 will at the same time expand the oxide layer and contract the metallic part of the cladding.
    b. If the oxide material zone and interface have already been created, the thickness of the cell layer at the oxide/metal interface is monitored, both on the oxide- and on the metal-side. Cell layers are added or removed depending on the threshold set by the user.

Overall, the explicit-oxide approach offers a more accurate modeling of cladding oxidation, in particular under extreme conditions, by explicitly modeling the additional oxide layer and its effect on thermal analysis and stress distribution, and by automatically detecting the metal/oxide interface location and temperature. However, the explicit-oxide approach's primary drawback is its computational complexity, which requires careful consideration of the mesh resolution to accurately capture thin oxide layers and prevent mesh distortion and poor quality. Also, creating and removing layers of cells requires selecting multiple appropriate addition and removal thresholds, and ensuring the continuity of stresses across different layer of materials is a well-known challenge. These factors must be carefully balanced with the simulation objectives to determine whether the additional accuracy provided by the explicit-oxide approach justifies the increased computational complexity compared to the implicit method, and whether the explicit-oxide approach is the best choice for a given application.

## 3. VERIFICATION

To verify the methodology described in the previous section, a simple test case is conducted using a 2-D axisymmetric fuel rod wedge irradiated for 1000 days. The linear heat generation rate is fixed at 10 kW/m, resulting in a heat flux of 298.9 W/m² at the outer cladding surface, and the outer oxide temperature is fixed at 600 K. Under these simplified conditions, the Richie model can be used to derive analytical values for the final oxide thickness and metal-oxide temperature.

Table 1 presents a comparison of the values predicted by OFFBEAT with the analytical counterpart for both the implicit- and explicit-oxide versions of the model. The table includes the interface temperature as well as the change in cladding thickness obtained from the model. The results demonstrate that the model implemented in OFFBEAT accurately predicts the final interface temperature and oxide layer thickness in both approaches, and that the ratio of oxide thickness to metal thickness $S_{ox}/S_{met}$ is equal to the Pilling-Bedworth ratio. Figure 2 shows a top view of the modeled geometry for both methodologies at 1000 days. As expected, the total final diameter of the cladding in the implicit approach is identical to the sum of the metallic cladding and oxide layer thickness calculated by the explicit approach. It was also verified that the deformation of the mesh points shown in the Figure corresponds to the values reported in the Table 1.

**Table I. Results of the verification test case**.

| Case | $T_I$ (Expected / Calculated) | $S_{ox}$ (Expected / Calculated) | Total $S_{met}$ | $S_{ox}/S_{met}$ |
|---|---|---|---|---|
| Implicit | 602.672 / 602.664 | 8.440 μm / 8.435 μm | 5.41 μm | 1.56 |
| Explicit | 602.672 / 602.668 | 8.440 μm / 8.435 μm | 5.41 μm | 1.56 |

## 4. TEST CASE

The aim of this section is to demonstrate the capabilities of our methodology to model the growth of non-uniform oxide layers. For instance, this can occur during dryout conditions in BWRs due to local power peaking and asymmetries in the temperature profile of the fuel rods. Such asymmetries, if sustained for an extended period, can result in the formation of relatively thick non-uniform oxide layers.

To demonstrate the potential of our approach, we simulate a 2-D cladding ring, using both our oxide modeling approaches. The computational mesh used for this case consists of 20 radial cells and 400 divisions around the whole azimuthal angle, and it is one cell thick in the axial direction. For simplicity, we do not include the fuel region, and we assume a constant and spatially uniform heat flux at the inner surface of the ring corresponding to a linear heat rate of 20 kW/m.

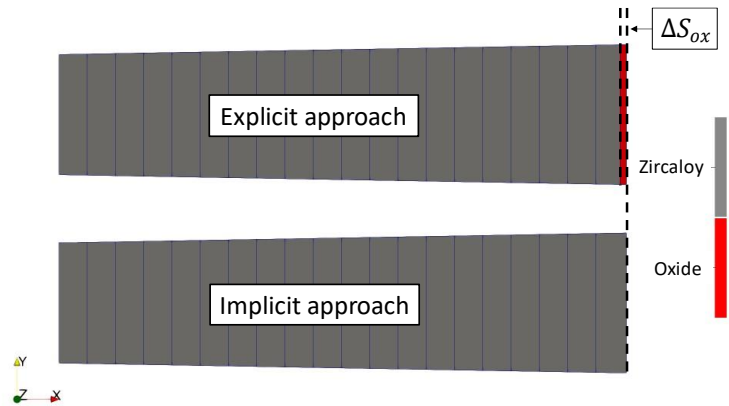

**Figure 2. Verification case – final mesh deformation with Explicit and Implicit approaches.**

To replicate realistic conditions representative of an asymmetric dryout, we reference the work of Clifford et al. [9] where the authors examined the causes of a rod leak in a Swiss BWR and analyzed the impact of local power gradients, pellet eccentricity, and neutronics feedback for various dryout angles. We use their findings to recreate an asymmetric outer temperature profile, based on the worst-case scenario identified in their paper, which corresponded to a dryout angle of 90 degrees at the EOC of cycle B (for reference, Figure 7 in the original paper). We apply the same temperature profile (which in the original paper was normalized to a non-specified nominal temperature) to a nominal temperature of 563 K, a typical saturated water temperature of a BWR. The profile reproduced in this work is shown on the left of Figure 3.

For our study, we assume the following simplified irradiation history: first, we simulate a base irradiation period with large time steps lasting for 3 years, during which the outer temperature on the cladding ring remains uniform and equal to 563K; then we change the time step size to 0.1 hrs to simulate a quasi-steady state transient that lasts for 24hrs. During this transient we assume that dryout condition occur with local power peaking and we apply the asymmetric temperature profile described above.

As a showcase of the capabilities of our methodology, in Figure 4 we show the final oxide layer predicted by the explicit approach: with the given mesh and with the chosen addition/removal thresholds and settings, the layer ended up being only one cell thick, although its thickness varies across the azimuthal angle. Upon closer examination, we found that the two approaches predict slightly different final thicknesses for the oxide layer: a maximum of ~36 μm for the implicit approach, and a maximum of ~34 μm for the explicit approach. Although this difference is minor, it is consistent with the difference in temperature profiles that we analyze in the next paragraph.

Indeed, to gain further insight into the thermal response predicted by the two models, we examine the temperature profiles at the end of the simulation in Figure 3. The figure displays the outer cladding temperature for the implicit approach, as well as the outer cladding (or outer oxide) temperature and the metal/oxide interface temperature for the explicit approach. As expected, the outer oxide temperature in the explicit approach matches the imposed asymmetric boundary condition. Similarly, away from the dryout region the explicit interface temperature is correctly higher than the outer temperature calculated with the implicit approach. However, we observe a reversal of this trend in the dryout region, where the outer temperature predicted by the implicit approach at its peak is ~10 degrees higher than the explicit interface temperature. We attribute this difference to the enhanced (2-D) thermal diffusion across the explicit oxide layer which is correctly considered only in the explicit approach. Returning to Figure 3, we now have an explanation for why the explicit approach predicts a slightly smaller oxide layer thickness: the presence of the oxide layer in the model contributes to lowering the interface temperature and leads to a slightly smaller final oxide thickness.

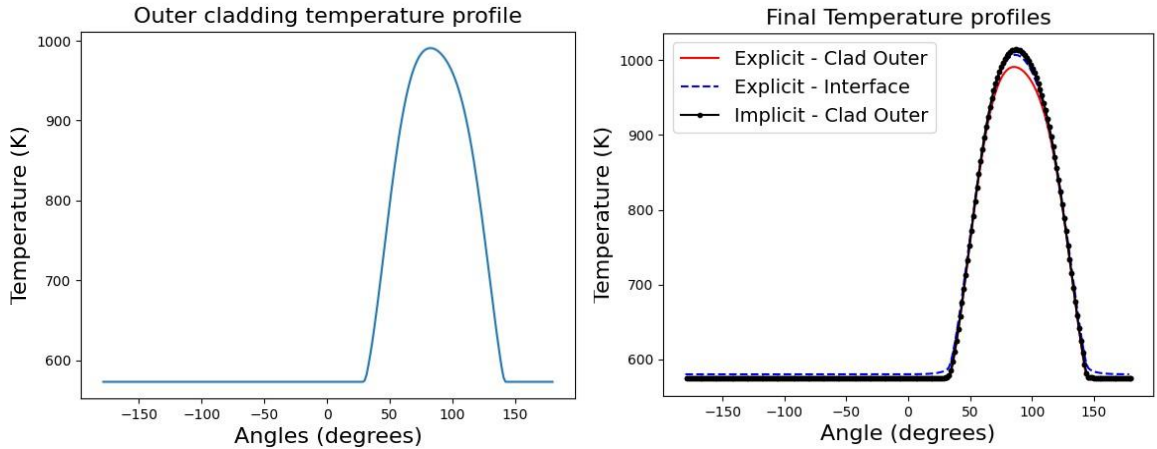

**Figure 3. Test case –Temperature profile prescribed at the boundary of the domain (left) and final temperature distributions for the explicit and implicit approaches (right).**

From the point of view of mechanical analysis, the discrepancies between the two approaches become more prominent. As an example, in Figure 5 we display the hoop stress profile in a similar way as it was done in Figure 4 with the temperature distribution. In the dryout region, the implicit outer stress falls between the outer and interface stresses predicted by the explicit approach. However, for the rest of the profile, there is a notable difference as the explicit approach shows a sharp decrease in the hoop stress (in magnitude) near the maximum dryout point, which is entirely absent from the implicit approach. This difference is attributed to the additional constraints deriving from an attached, multi-layered domain, which are absent from the implicit approach. It can be assumed that similar discrepancies will increase further as the oxide layer becomes thicker.

## 5. CONCLUSIONS

In this work, we have presented the first steps toward the development of a methodology for modeling the formation of non-uniform oxide layers in Zircaloy claddings and studying their impact on the rod's thermal and mechanical response under irradiation. Our approach offers two possible alternatives: an implicit-oxide method that preserves the total cross-sectional area of the corroded cladding but does not account for the oxide layer with separate mesh cells; and an explicit-oxide method which tracks the growth of the oxide layer with a separate material zone while also decreasing the thickness of the metallic portion of the cladding.

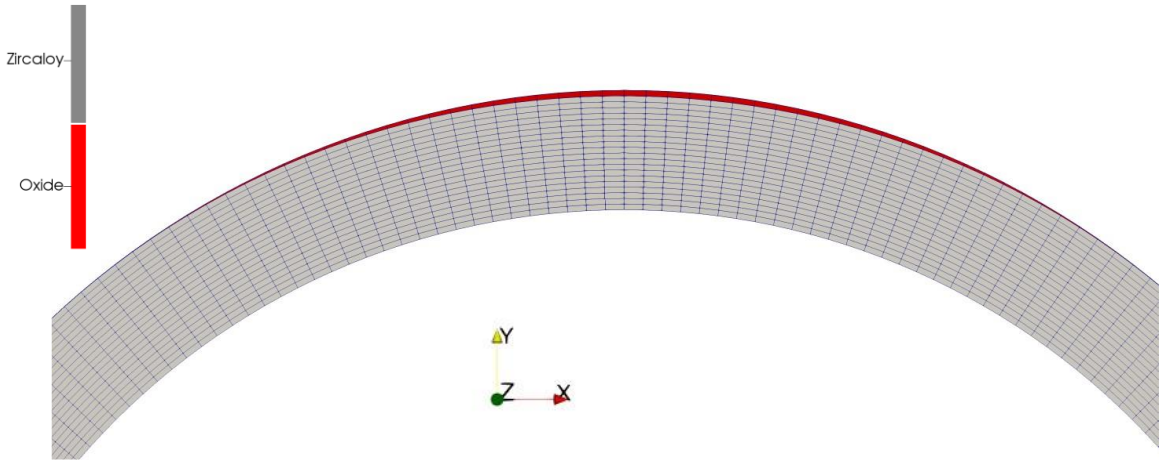

**Figure 4. Test-case – Top view of the layer created by the explicit approach.**

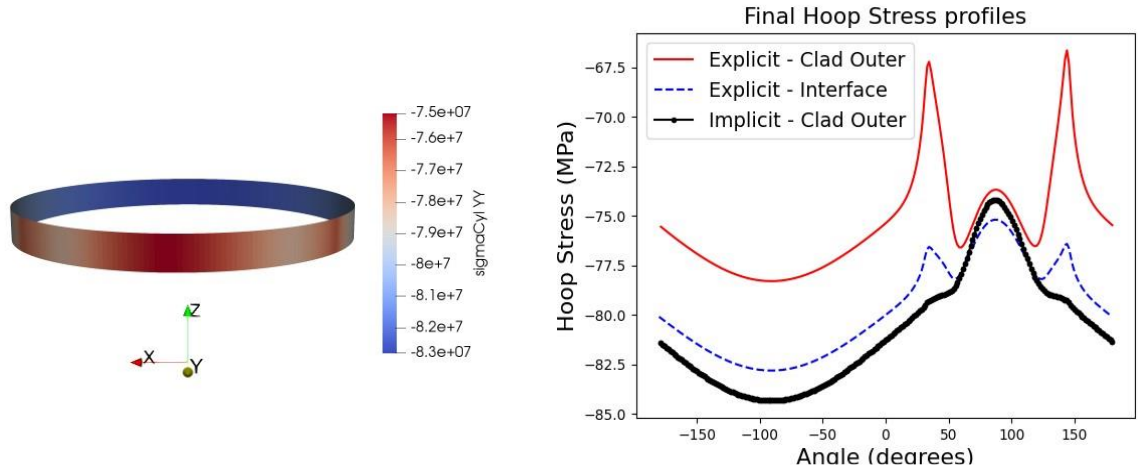

**Figure 5. Test-case – Final hoop stress distribution for the implicit and explicit approaches.**

Both approaches were implemented in the OFFBEAT fuel performance code and were verified against a simple test case showing good agreement with the analytical solution. The capabilities of the methodology to deal with non-uniform oxide layers were then demonstrated by studying a representative cladding ring mimicking local power peaking conditions during dryouts in BWRs.

Future works should focus on improving the stability of the layer addition/removal algorithm for complex asymmetries and on accurately verifying the stress distribution at the interface in the case of the explicit approach. Also, validation against experimental data or axisymmetric test cases available in the open literature is needed to boost confidence in the methodology. Also, more efforts should be dedicated to clearly identifying cases where the differences between the implicit and explicit approaches become significant.

Possible future applications of the methodology are numerous. For instance, it would be interesting to model a power peaking scenario like the one discussed in this paper, but this time modeling the full irradiation history for a rodlet or for the full-length rod, as well as incorporating the coupling with CFD for more realistic boundary conditions. A similar combination with CFD simulations might also be used to study the effect of fuel assembly spacer grids on the rod-to-coolant heat transfer and thus on the axial variation of local oxide thickness.

Furthermore, the mesh modification methodology presented in this work offers potential applications beyond corrosion itself. With the implementation of few additional features already available in OpenFOAM, it could be extended to study phenomena such as crud deposition, the detachment/attachment of TRISO layers, the formation of the JOG in MOX type fuels or the formation of layers of separate materials in plate type fuel.

In summary, our methodology provides a promising approach to studying non-uniform oxide layer growth in metallic fuel cladding, with potential interesting applications in the field of nuclear simulations. With further development and validation, it could become a valuable tool for predicting the performance and safety of nuclear fuel rods.

## ACKNOWLEDGMENTS

The authors would like to acknowledge the support of the IAEA ONCORE initiative on open-source software.